%% file: AsyGamArchive.tex
\documentclass[10pt]{article}

\usepackage{amssymb,amsmath,amsthm,mathrsfs}
\usepackage[dvips]{graphics,epsfig}
\usepackage{verbatim}
\usepackage{colordvi,color}
\usepackage{psfrag}
\usepackage{colordvi,color,pspicture}
\usepackage{url}

\graphicspath{{Figures/}}

\newcommand{\Y}{\mathcal{Y}}
\newcommand{\X}{\mathcal{X}}
\newcommand{\NY}{N_{\mathcal{Y}}}
\newcommand{\G}{\Gamma}
\newcommand{\g}{\gamma}
\newcommand{\R}{\mathbb{R}}

\newcommand{\V}{\mathcal{V}}
\newcommand{\A}{\mathcal{A}}
\newcommand{\y}{\mathsf{y}}
\newcommand{\ve}{\varepsilon}
\newcommand{\U}{\mathcal{U}}

\newcommand{\E}[1]{\mathbf{E}\,#1}
\newcommand{\Var}[1]{\mathbf{Var}\,#1}

\DeclareMathOperator{\argmin}{argmin}
\DeclareMathOperator{\argmax}{argmax}
 
\DeclareMathOperator{\support}{support}

\begin{document}
\title{The Use of Domination Number of a Random Proximity Catch
Digraph for Testing Spatial Patterns of Segregation and Association$^\star$}
\author{\it Elvan Ceyhan \& Carey E. Priebe\\
\it Johns Hopkins University, Baltimore}
\date{\today}

\maketitle
\begin{abstract}
Priebe et al. (2001) introduced the class cover
catch digraphs and computed the distribution of the domination
number of such digraphs for one dimensional data. In higher
dimensions these calculations are extremely difficult due to the
geometry of the proximity regions; and only upper-bounds are
available. In this article, we introduce a new type of data-random
proximity map and the associated (di)graph in $\mathbb R^d$. We find
the asymptotic distribution of the domination number
and use it for testing spatial point patterns of segregation and association.
\end{abstract}

\vspace{0.1 in}


\vspace{3.0 in}

$^\star$This research was supported by the Defense Advanced
Research Projects Agency as administered by the Air Force Office of
Scientific Research under contract DOD F49620-99-1-0213 and by
Office of Naval Research Grant N00014-95-1-0777.\\
\indent
Corresponding author.\\
\indent {\it E-mail address:} cep@jhu.edu (C.E.~Priebe)

\section{Introduction}
In a digraph $D=(\V,\A)$ with vertex set $\V$ and arc (directed
edge) set $\A$, a vertex $v$ {\em dominates} itself and all vertices
of the form $\{u:\,vu \in \A\}$. A {\em dominating set}, $S_D$, for
the digraph $D$ is a subset of $\V$ such that each vertex $v \in \V$
is dominated by a vertex in $S_D$. A {\em minimum dominating set},
$S_D^*$, is a dominating set of minimum cardinality; and the {\em
domination number}, $\g(D)$, is defined as $\g(D):=|S_D^*|$,
 where $|\cdot|$ is the cardinality functional (\cite{west:2001}).
If a minimum dominating set is of size one, we call it a {\em
dominating point}.

Let $(\Omega,\mathcal{M})$ be a measurable space and consider a
function $N:\Omega \times 2^\Omega \rightarrow 2^\Omega$, where
$2^\Omega$ represents the power set of $\Omega$. Then given $\Y
\subseteq \Omega$, the {\em proximity map} $\NY(\cdot) =
N(\cdot,\Y): \Omega \rightarrow 2^\Omega$ associates with each point
$x \in \Omega$ a {\em proximity region} $\NY(x) \subset \Omega$. The
region $\NY(x)$ depends on the distance between $x$ and $\Y$. For $B
\subseteq \Omega$, the {\em $\G_1$-region},
$\G_1(\cdot)=\G_1(\cdot,\NY):\Omega \rightarrow 2^\Omega$ associates
the region $\G_1(B):=\{z \in \Omega: B \subseteq  \NY(z)\}$ with
each set $B \subseteq \Omega$. For $x \in \Omega$, we denote
$\G_1(\{x\})$ as $\G_1(x)$.

If $\X_n=\{X_1,X_2,\cdots,X_n\}$ is a set of $\Omega$-valued random
variables, then the $\NY(X_i)$ (and $\G_1(X_i)$), $i=1,\cdots,n$ are
random sets. If the $X_i$ are independent and identically
distributed, then so are the random sets $\NY(X_i)$ (and
$\G_1(X_i)$). Furthermore, $\G_1(\X_n)$ is a random set. Notice that
$\G_1(\X_n)=\cap_{j=1}^n \G_1(X_j)$, since $x \in \G_1(\X_n)$ iff
$\X_n \subseteq \NY(x)$ iff $X_j \in \NY(x)$ for all $j=1,\ldots,n$
iff $x \in \G_1(X_j)$ for all $j=1,\ldots,n$ iff $x \in \cap_{j=1}^n
\G_1(X_j)$.

Consider the data-random proximity catch digraph $D$ with vertex set
$\V=\X_n$ and arc set $\A$ defined by $(X_i,X_j) \in \A \iff X_j \in
\NY(X_i)$. The random digraph $D$ depends on the (joint)
distribution of the $X_i$ and on the map $\NY$ (see Priebe et al.
(2001) and Priebe et al. (2003)). The adjective {\em proximity}
--- for the catch digraph $D$ and for the map $\NY$ ---
comes from thinking of the region $\NY(x)$ as representing those
points in $\Omega$ ``close'' to $x$ (see, e.g., Toussaint (1980) and
Jaromczyk and Toussaint (1992)).

For $X_1,\cdots,X_n \stackrel{iid}{\sim} F$ the domination number of
the associated data-random proximity catch digraph $D$, denoted
$\g(\X_n;F,\NY)$, is the minimum number of points that dominate all
points in $\X_n$. Note that, $\g(\X_n;F,\NY)=1$ iff $\X_n \cap
\G_1(\X_n)\not= \emptyset$.

The random variable $\g_n := \g(\X_n;F,\NY)$ depends on $\X_n$
explicitly, and on $F$ and $\NY$ implicitly. In general, the
expectation $\E[\g_n]$, depends on $n$, $F$, and $\NY$; $ 1 \leq
\E[\g_n] \le n$; and the variance of $\g_n$ satisfies, $0\le
\Var[\g_n] \le n^2/4$.

We can also define the regions associated with $\g_n=k$ for $k \le
n$. For instance, the $\G_2$-region for proximity map
$N_{\Y}(\cdot)$ and set $B \subset \Omega$ is $\G_2(B)=\{(x,y) \in
[\Omega \setminus \G_1(B)]^2:\;B \subseteq N_{\Y}(x)\cup
N_{\Y}(y)\}$. In general,
\begin{multline*}
\G_k(B)=\{(x_1,x_2,\ldots,x_k) \in \Omega^k: B \subseteq \cup_{j=1}^k N_{\Y}(x_j) \;\text{ and all possible $m$-permutations $(u_1,u_2,\ldots,u_m)$}\\
 \text{ of $(x_1,x_2,\ldots,x_k)$
 satisfy }(u_1,u_2,\ldots,u_m) \not\in \G_m(B) \text{ for each } m=1,2,\ldots,k-1 \}.
\end{multline*}

\section{A Class of Proximity Maps and the Corresponding $\G_1$-Regions}
Let $\Omega = \R^2$ and let $\Y = \{\y_1,\y_2,\y_3\} \subset \R^2$
be three non-collinear points. Denote by $T(\Y)$ the triangle
---including the interior--- formed by these three points. The most
straightforward extension of the data random proximity catch digraph
introduced by Priebe et al. (2001) is the spherical proximity map
$N_S(x)=B(x,r(x))$ which is the ball centered at $x$ with radius
$r(x)=\min_{\y \in \Y}d(x,\y)$ or the arc-slice proximity map
$N_{AS}(x)=B(x,r(x)) \cap T(\Y)$. However, both cases suffer from
the intractability of the $\G_1$-region and hence the intractability
of the finite and asymptotic distribution of $\g_n$. We propose a
new class of proximity regions which does not suffer from this
drawback.

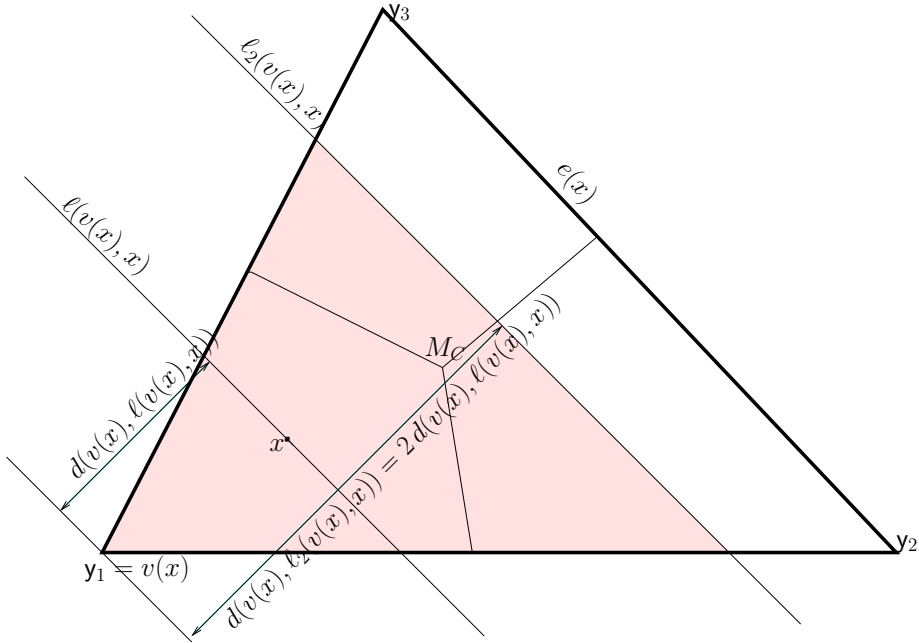
\begin{figure} [ht]
    \centering
   \scalebox{.5}{\input{Nofnu2.pstex_t}}
   \caption{Construction of $r$-factor proximity region, $\NY^2(x)$ (shaded region).} 

\label{fig:ProxMapDef1}
    \end{figure}

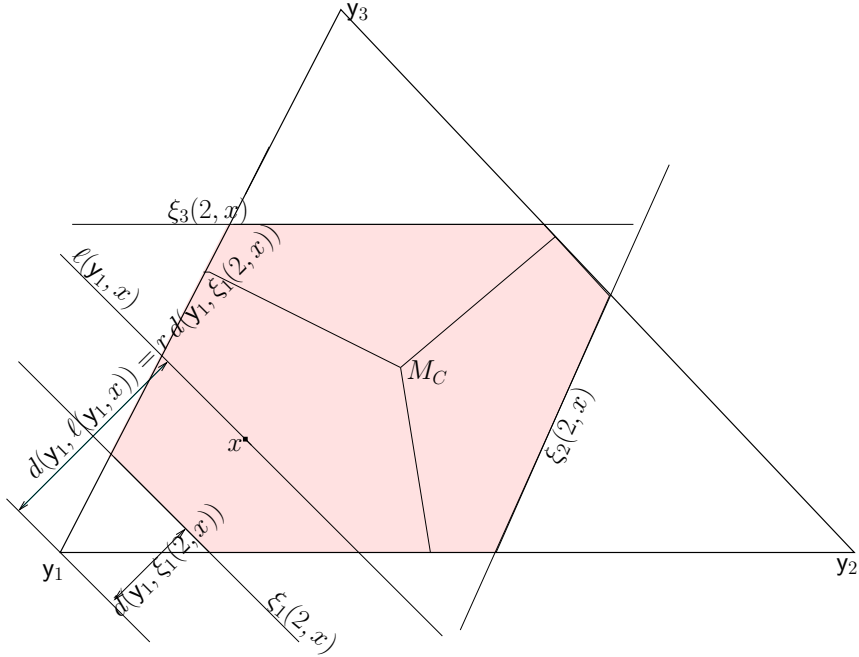
\begin{figure} [ht]
    \centering
    \scalebox{.5}{\input{Gammaofnu.pstex_t}}
    \caption{Construction of the $\G_1$-region, $\G_1^2(x)$ (shaded region). }
\label{fig:ProxMapDef2}
    \end{figure}

For $r \in [1,\infty]$ define $\NY^r$ to be the {\em r-factor
proximity map} and $\G_1^r$ to be the corresponding $\G_1$-region as
follows; see also Figures  \ref{fig:ProxMapDef1} and
\ref{fig:ProxMapDef2}. Let ``vertex regions'' $R(\y_1)$, $R(\y_2)$,
$R(\y_3)$ partition $T(\Y)$ using segments from the center of mass
of $T(\Y)$ to the edge midpoints. For $x \in T(\Y) \setminus \Y$,
let $v(x) \in \Y$ be the vertex whose region contains $x$; $x \in
R(v(x))$. If $x$ falls on the boundary of two vertex regions, we
assign $v(x)$ arbitrarily. Let $e(x)$ be the edge of $T(\Y)$
opposite $v(x)$. Let $\ell(v(x),x)$ be the line parallel to $e(x)$
through $x$. Let $d(v(x),\ell(v(x),x))$ be the Euclidean
(perpendicular) distance from $v(x)$ to $\ell(v(x),x)$. For $r \in
[1,\infty)$ let $\ell_r(v(x),x)$ be the line parallel to $e(x)$ such
that $d(v(x),\ell_r(v(x),x)) = rd(v(x),\ell(v(x),x))$ and
$d(\ell(v(x),x),\ell_r(v(x),x)) < d(v(x),\ell_r(v(x),x))$. Let
$T_r(x)$ be the triangle similar to and with the same orientation as
$T(\Y)$ having $v(x)$ as a vertex and $\ell_r(v(x),x)$ as the
opposite edge. Then the {\em r-factor} proximity region $\NY^r(x)$
is defined to be $T_r(x) \cap T(\Y)$.

To define the $\G_1$-region, let $\xi_j(x)$ be the line such that
$\xi_j(x)\cap T(\Y) \not=\emptyset$ and
$r\,d(\y_j,\xi_j(x))=d(\y_j,\ell(\y_j,x))$
 for $j=1,2,3$.  Then $\G_1^r(x)=\cup_{j=1}^3 \bigl(\G_1^r(x)\cap R(\y_j) \bigr)$ where
$\G_1^r(x)\cap R(\y_j)=\{z \in R(\y_j): d(\y_j,\ell(\y_j,z)) \ge
d(\y_j,\xi_j(x)\}$, for $j=1,2,3$. Notice that $r \ge 1$ implies $x
\in \NY^r(x)$ and $x \in \G_1^r(x)$. Furthermore, $\lim_{r
\rightarrow \infty} \NY^r(x) = T(\Y)$ and  $\lim_{r \rightarrow
\infty} \G_1^r(x) = T(\Y)$ for all $x \in T(\Y) \setminus \Y$, and
so we define $\NY^{\infty}(x) = T(\Y)$ and $\G_1^{\infty}(x) =
T(\Y)$ for all such $x$. For $x \in \Y$, we define $\NY^r(x) =
\{x\}$ for all $r \in [1,\infty]$.

Notice that $X_i \stackrel{iid}{\sim} F$, with the additional
assumption that the non-degenerate two-dimensional probability
density function $f$ exists with $\support(f) \subseteq T(\Y)$,
implies that the special case in the construction of $\NY^r$ --- $X$
falls on the boundary of two vertex regions --- occurs with
probability zero. Note that for such an $F$, $\NY^r(x)$ is a
triangle a.s. and $\G^r_1(x)$ is a star-shaped polygon (not
necessarily convex).

Let $X_e:=\argmin_{X \in \X_n}d(X,e)$ be the (closest) edge extremum
for edge $e$. Then $\G^r_1(\X_n)=\cap_{j=1}^3\G^r_1(X_{e_j})$, where
$e_j$ is the edge opposite vertex $\y_j$, for $j=1,2,3$. So
$\G^r_1(\X_n)\cap R(\y_j)=\{z \in R(\y_j):\; d(\y_j, \ell(\y_j,z)
\ge d(\y_j,\xi_j(X_{e_j}))\}$, for $j=1,2,3$.

Let the domination number be $\g_n(r):=\g_n(\X_n;F,\NY^r)$ and
$X_{[j]}:=\argmin_{X\in \X_n \cap R(\y_j)}d(X,e_j)$. Then $\g_n(r)
\le 3$ with probability 1, since $\X_n \cap R(\y_j) \subset
\NY^r(X_{[j]})$ for each $j=1,2,3$. Thus
$$1 \le \E[\g_n(r)] \le 3 \text{    and    } 0 \le \Var[\g_n(r)] \le 9/4.$$

\section{Null Distribution of Domination Number}
The null hypothesis for spatial patterns have been a contraversial
topic in ecology from the early days. \cite{gotelli:1996} have
collected a voluminous literature to present a comprehensive
analysis of the use and misuse of null models in ecology community.
They also define and attempt to clarify the null model concept as
``a pattern-generating model that is based on randomization of
ecological data or random sampling from a known or imagined
distribution. . . . The randomization is designed to produce a
pattern that would be expected in the absence of a particular
ecological mechanism." In other words, the hypothesized null models
can be viewed as ``thought experiments," which is conventially used
in the physical sciences, and these models provide a statistical
baseline for the analysis of the patterns. For statistical testing,
the null hypothesis we consider is a type of {\em complete spatial
randomness}; that is,
$$H_0: X_i \stackrel{iid}{\sim} \U(T(\Y))$$
where $\U(T(\Y))$ is the uniform distribution on $T(\Y)$. If it is
desired to have the sample size be a random variable, we may
consider a spatial Poisson point process on $T(\Y)$ as our null
hypothesis.

We first present a ``geometry invariance" result which allows us to
assume $T(\Y)$ is the standard equilateral triangle,
$T\bigl((0,0),(1,0),\bigl( 1/2,\sqrt{3}/2 \bigr)\bigr)$, thereby
simplifying our subsequent analysis.

{\bf Theorem 1}: Let $\Y = \{\y_1,\y_2,\y_3\} \subset \R^2$ be three
non-collinear points. For $i=1,\cdots,n$, let $X_i
\stackrel{iid}{\sim} F = \U(T(\Y))$, the uniform distribution on the
triangle $T(\Y)$. Then for any $r \in [1,\infty]$ the distribution
of $\g(\X_n;\U(T(\Y)),\NY^r)$ is independent of $\Y$, and hence the
geometry of $T(\Y)$.

{\bf Proof:} A composition of translation, rotation, reflections,
and scaling will take any given triangle $T_o =
T(\{\y_1,\y_2,\y_3\})$ to the ``basic'' triangle $T_b =
T(\{(0,0),(1,0),(c_1,c_2)\})$ with $0 < c_1 \le 1/2$, $c_2 > 0$ and
$(1-c_1)^2+c_2^2 \le 1$, preserving uniformity. The transformation
$\phi_e: \R^2 \rightarrow \R^2$ given by $\phi_e(u,v) = \left(
u+\frac{1-2\,c_1}{\sqrt{3}}\,v,\frac{\sqrt{3}}{2\,c_2}\,v \right)$
takes $T_b$ to the equilateral triangle $T_e =
T(\{(0,0),(1,0),(1/2,\sqrt{3}/2)\})$. Investigation of the Jacobian
shows that $\phi_e$ also preserves uniformity. Furthermore, the
composition of $\phi_e$ with the rigid motion transformations maps
     the boundary of the original triangle, $T_o$,
  to the boundary of the equilateral triangle, $T_e$,
     the median lines of $T_o$
  to the median lines of $T_e$,
and  lines parallel to the edges of $T_o$
  to lines parallel to the edges of $T_e$.
Since the distribution of $\g(\X_n;\U(T(\Y)),\NY^r)$ involves only
probability content of unions and intersections of regions bounded
by precisely such lines, and the probability content of such regions
is preserved since uniformity is preserved, the desired result
follows. $\blacksquare$

Based on Theorem 1 and our uniform null hypothesis, we may assume
that $T(\Y)$ is a standard equilateral triangle with $\Y =
\{(0,0),(1,0),(1/2,\sqrt{3}/2)\}$ henceforth.

For our $r$-factor proximity map and uniform null hypothesis, the
asymptotic null distribution of $\g_n(r) :=
\g(\X_n;\U(T(\Y)),\NY^r)$
 can be derived as a function of $r$.
We denote by $\varsigma^r_{\Y}:=\{z \in T(\Y):\;\NY^r(z)=T(\Y)\}$
the {\em superset region} associated with $\NY^r$ in $T(\Y)$. Notice
that $\varsigma^r_{\Y} \subseteq \G_1^r(\X_n)$ for all $r$ and $\X_n
\cap \varsigma^r_{\Y}\not= \emptyset$ implies that $\g_n(r)=1$.

{\bf Proposition 1:} The expected area of the the $\G_1$-region,
$\E[A(\G^r_1(\X_n))]$, converges to the area of the superset region,
$A(\varsigma^r_{\Y})$, as $n \rightarrow \infty$. In particular,
$\E[A(\G^{3/2}_1(\X_n))]$, goes to zero at rate $O(n^{-2})$ as $n
\rightarrow \infty$.

{\bf Proof:} See Appendix. $\blacksquare$

As a corollary to the above proposition, we have that
$\E[A(\G^r_1(\X_n))] \rightarrow A(\varsigma^r_{\Y})=0$ for $r \in
[1,3/2]$ as $n \rightarrow \infty$. Additionally,
$\E[A(\G^r_1(\X_n))] \rightarrow
A(\varsigma^r_{\Y})=(1-3/(2\,r))^2\,\sqrt{3}$ for $ r \in (3/2,2]$,
and $\E[A(\G^r_1(\X_n))] \rightarrow
A(\varsigma^r_{\Y})=\sqrt{3}/4\,(1-3/r^2)$ for $ r \in (2,\infty]$,
as $n \rightarrow \infty$.

{\bf Theorem 2:} The domination number $\g_n(r) =
\g(\X_n;\U(T(\Y)),\NY^r)$ is degenerate in the limit for $r \in
[1,\infty]\setminus \{3/2\}$ as $n \rightarrow \infty$.

{\bf Proof:} For $r \in [1,3/2)$, $\varsigma^r_{\Y}=\emptyset$ and
$T(\Y)\setminus \NY^r(X_{[j]})$ has positive area for all $j=1,2,3$.
Furthermore, $T(\Y)\setminus (\NY^r(X_{[j]})\cup \NY^r(X_{[k]}))$
has positive area for all pairs $\{j,k\} \subset \{1,2,3\}$. Recall
that $\g_n(r) \le 3$ with probability 1 for all $n$ and $r$. Hence
$\g_n(r) \rightarrow 3$ in probability as $n \rightarrow \infty$.

For $r \in (3/2,\infty]$, $\varsigma^r_{\Y}$ has positive area, so
$\g_n(r) \rightarrow 1$ in probability as $n \rightarrow \infty$.
$\blacksquare$

{\bf Theorem 3:} For $r=3/2$, $\lim_{n \rightarrow \infty}\g_n(r) >
1$ a.s. In particular
\begin{eqnarray*}
\label{eq:Asydist} \lim_{n \rightarrow \infty} \g_n(3/2) =
 \begin{cases}
   2  &\text{wp $\approx$} \quad .7413, \\
   3  &\text{wp $\approx$} \quad .2487.
 \end{cases}
\end{eqnarray*}
Thus $\E[\g_n(3/2)]\rightarrow \mu \approx 2.2587$ as $n\rightarrow
\infty$, and $\Var[\g_n(3/2)]\rightarrow \sigma^2 \approx .1918$ as
$n\rightarrow \infty$.

{\bf Proof:} See Appendix. $\blacksquare$

\begin{figure}[ht]
\centering \rotatebox{-90}{ \resizebox{2.5 in}{!}{
\includegraphics{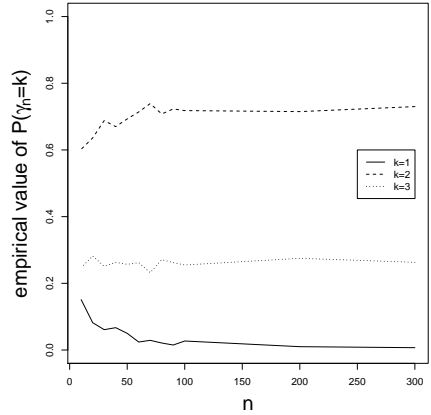} } } \caption{Plotted are the
empirical estimates of $P(\g_n(3/2)=k)$ versus various $n$ values.}
\label{fig:Emp pdf}
\end{figure}

The finite sample distribution of $\g_n(3/2)$, and hence the finite
sample mean and variance, can be obtained by numerical methods. We
estimate the distribution of $\g_n(3/2)$ for various fixed $n$
empirically. In Table \ref{tab:emp-dist}, we present empirical
estimates for $n=10,20,\ldots,100,200,300$ with $1000$ Monte Carlo
replicates. See also Figure \ref{fig:Emp pdf}.

\begin{table}[t]
\centering
\begin{tabular}{|c|c|c|c|c|c|c|c|c|c|c|c|c|}
\hline
$ k \diagdown n$  & 10 & 20 & 30 & 40 & 50 & 60 & 70 & 80 & 90 & 100 & 200 & 300\\
\hline
1  & 151 & 82 & 61 & 67 & 50 & 24 & 29 & 21 & 15 & 27 & 10 & 7\\
\hline
2 &  602 & 636 & 688 & 670 & 693 & 714 &  739 & 708 & 723 & 718 & 715 & 730 \\
\hline
3  & 247 & 282 & 251 & 263 & 257 & 262 & 232 & 271 & 262 & 255 & 275 & 263\\
\hline
\end{tabular}
\caption{ \label{tab:emp-dist} The number of $\g_n(3/2)=k$ out of
$N=1000$ replicates.}
\end{table}

{\bf Theorem 4} Let $\g_n(r) = \g(\X_n;\U(T(\Y)),\NY^r)$.  Then
$r_1<r_2$ implies $\g_n(r_2) <^{ST} \g_n(r_1)$.

{\bf Proof:} Suppose $r_1<r_2$.  Then
$P(\g_n(r_2)=1)>P(\g_n(r_1)=1)$  and $P(\g_n(r_2)=2)>P(\g_n(r_1)=2)$
and $P(\g_n(r_2)=3)<P(\g_n(r_1)=3)$. Hence the desired result
follows. $\blacksquare$

\section{The Null Distribution of the Mean Domination Number in the Multiple Triangle Case}

Suppose $\Y$ is a finite collection of points in $\R^2$ with $|\Y|
\ge 3$. Consider the Delaunay triangulation (assumed to exist) of
$\Y$, where $T_j$ denotes the $j^{th}$ Delaunay triangle, $J$
denotes the number of triangles, and $C_H(\Y)$ denotes the convex
hull of $\Y$ (Okabe et al. (2000)). We wish to investigate
$$H_0: X_i \stackrel{iid}{\sim} \U(C_H(\Y))$$
against segregation and association alternatives (see Section
\ref{sec:alternatives}).

\begin{figure}[ht]
\centering \epsfig{figure=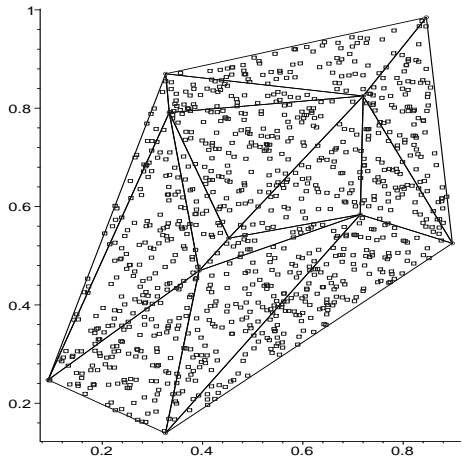, height=170pt,
width=170pt} \caption{ \label{fig:deldatacsr} A realization of $H_0$
for $|\Y|=10$, $J=13$, and $n=1000$. }
\end{figure}

Figure \ref{fig:deldatacsr} presents a realization of 1000
observations independent and identically distributed according to
$\U(C_H(\Y))$ for $|\Y|=10$ and $J=13$.

The digraph $D$ is constructed using $N_{\Y_j}^r(\cdot)$ as
described above, where for $X_i \in T_j$ the three points in $\Y$
defining the Delaunay triangle $T_j$ are used as $\Y_j$. Let
$\g_{n_j}(r)$ be the domination number of the component of the
digraph in $T_j$, where $n_j=|\X_n \cap T_j|$.

{\bf Theorem 5:} (Asymptotic Normality) Suppose $n_j \gg 1$ and $J$
is sufficiently large.  Then the null distribution of the mean
domination number $\overline G_J:=\frac{1}{J}\,\sum_{j=1}^J
\g_{n_j}(3/2)$ is given by
$$ \overline G_J \stackrel {\text{approx}}{\sim} \mathcal N (\mu,\sigma^2/J)$$
where $\mu$ and $\sigma^2$ are given in Theorem 3 above.

{\bf Proof:} For fixed $J$ sufficiently large and each $n_j$
sufficiently large, $\g_{n_j}(3/2)$ are approximately independent
identically distributed as in Theorem 3. $\blacksquare$

\begin{figure}[ht]
\centering \psfrag{Density}{ \Huge{\bf{Density}}} \rotatebox{-90}{
\resizebox{2.1 in}{!}{ \includegraphics{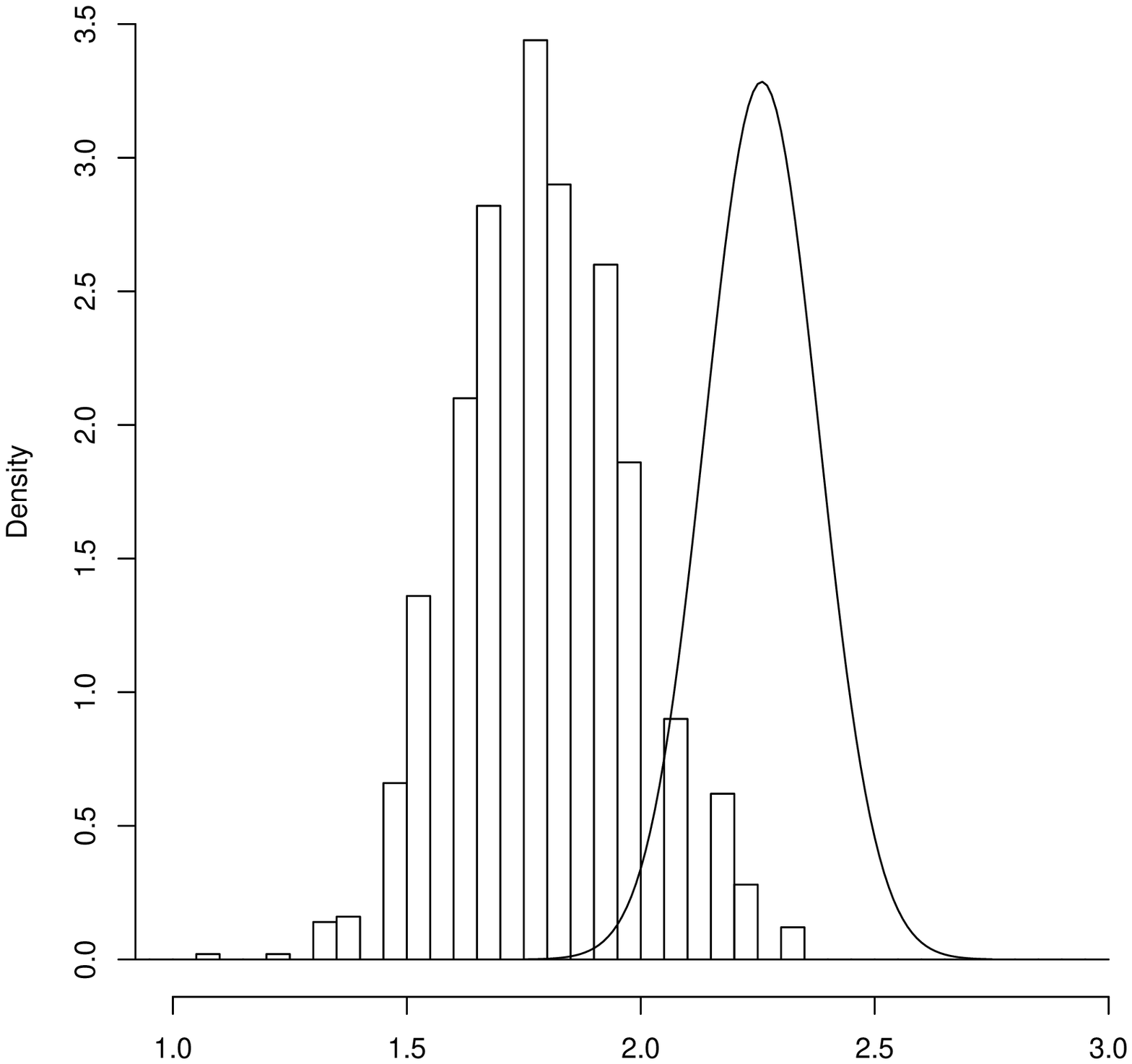} } }
\rotatebox{-90}{ \resizebox{2.1 in}{!}{
\includegraphics{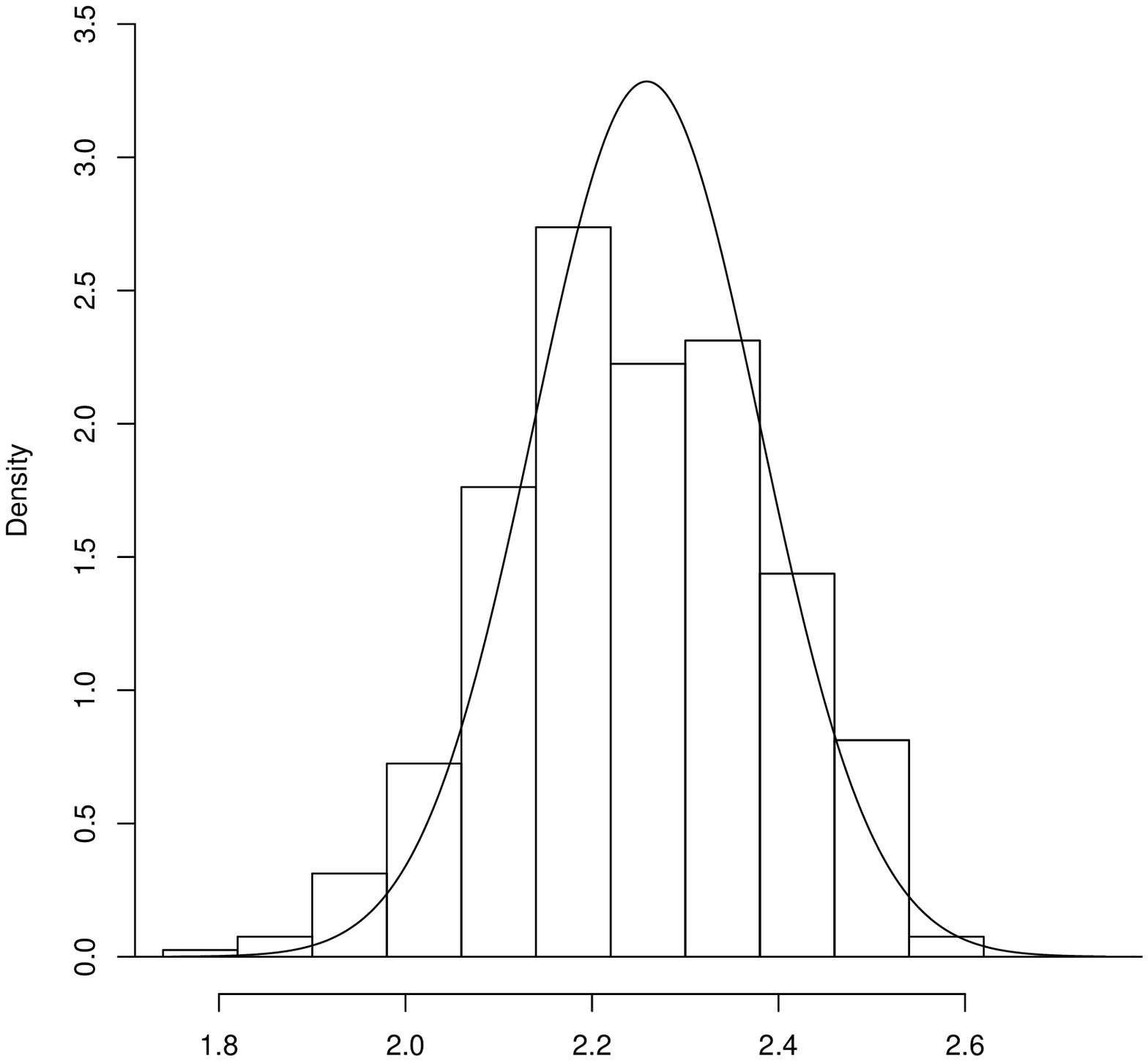} } }
\caption{ \label{fig:NormSkew} Depicted are $\overline G_J
\stackrel{\text{approx}}{\sim} \mathcal{N}(\mu \approx
2.2587,\sigma^2/J \approx.1917/J )$ for $J=13$ and $n=100$ (left)
$n=1000$ (right). Histograms are based on $1000$ Monte Carlo
replicates and the curves are the associated approximating normal
curves. }
\end{figure}

Figure \ref{fig:NormSkew}
indicates that, for $J=13$ with the realization of $\Y$ given in Figure \ref{fig:deldatacsr} and $n=100$ the normal approximation is not appropriate, even though the distribution looks symmetric, since not all $n_j$ are sufficiently large, but for $n=1000$ the histogram and the corresponding normal curve are similar indicating that this sample size is large enough to allow the use of the asymptotic normal approximation, since all $n_j$ are sufficiently large.  However, larger $J$ values require larger sample sizes in order to obtain approximate normality. 

For finite $n$, let $\overline{G}_J(r)$ be the mean domination
number associated with the digraph based on $N^r_{\Y}$.  Then as a
corollary to Theorem 4 it follows that for $r_1<r_2$,  we have
$\overline{G}_J(r_2) <^{ST}\overline{G}_J(r_1)$.

\section{Alternatives: Segregation and Association}
\label{sec:alternatives} In a two class setting, the phenomenon
known as {\em segregation} occurs when members of one class have a
tendency to repel members of the other class. For instance, it may
be the case that one type of plant does not grow well in the
vicinity of another type of plant, and vice versa. This implies, in
our notation, that $X_i$ are unlikely to be located near any
elements of $\Y$. Alternatively, association occurs when members of
one class have a tendency to attract members of the other class, as
in symbiotic species, so that the $X_i$ will tend to cluster around
the elements of $\Y$, for example. See, for instance,
\cite{dixon:1994}, \cite{coomes:1999}.

We define two simple classes of alternatives, $H^S_{\ve}$ and
$H^A_{\ve}$ with $\ve \in (0,\sqrt{3}/3)$, for segregation and
association, respectively. Let
$\Y_e=\{(0,0),(0,1),(1/2,\sqrt{3}/2)\}$ and $T_e=T(\Y_e)$. For $\y
\in \Y_e$, let $e(\y)$ denote the edge of $T_e$ opposite vertex
$\y$, and for $x \in T_e$ let $\ell_{\y}(x)$ denote the line
parallel to $e(\y)$ through $x$. Then define $T(\y,\ve) = \{x \in
T_e: d(\y,\ell_{\y}(x)) \le \ve\}$. Let $H^S_{\ve}$ be the model
under which $X_i \stackrel{iid}{\sim} \U(T_e \setminus \cup_{\y \in
\Y} T(\y,\ve))$ and $H^A_{\ve}$ be the model under which $X_i
\stackrel{iid}{\sim} \U(\cup_{\y \in \Y} T(\y,\sqrt{3}/3 - \ve))$.
Thus the segregation model excludes the possibility of any $X_i$
occurring near a $\y_j$, and the association model requires that all
$X_i$ occur near $\y_j$'s. The $\sqrt{3}/3 - \ve$ in the definition
of the association alternative is so that $\ve=0$ yields $H_0$ under
both classes of alternatives.

{\bf Remark:} These definitions of the alternatives are given for
the standard equilateral triangle. The geometry invariance result of
Theorem 1 still holds under the alternatives, in the following
sense. If, in an arbitrary triangle, a small percentage $\delta
\cdot 100\%$ where $\delta \in (0,4/9)$ of the area is carved away
as forbidden from each vertex using line segments parallel to the
opposite edge, then under the transformation to the standard
equilateral triangle this will result in the alternative
$H^S_{\sqrt{3 \delta / 4}}$. This argument is for segregation; a
similar construction is available for association.

{\bf Theorem 6:} (Stochastic Ordering) Let $\g_{n,\ve}(r)$ be the
domination number under the segregation alternative with $\ve>0$.
Then with $\ve_j \in (0,\sqrt{3}/3)$, $j=1,2$, $\ve_1 > \ve_2$
implies that $\g_{n,\ve_1}(3/2) <^{ST} \g_{n,\ve_2}(3/2)$.

{\bf Proof:} Note that
$P(\g_{n,\ve_1}(3/2)=1)>P(\g_{n,\ve_2}(3/2)=1)$ and
$P(\g_{n,\ve_1}(3/2)=2)>P(\g_{n,\ve_2}(3/2)=2)$, hence the desired
result follows. $\blacksquare$

Note that for Theorem 6 to hold in the limiting case, $\ve_1 \in
(0,\sqrt{3}/4]$ and $\ve_2 \in (\sqrt{3}/4,\sqrt{3}/3)$ should hold.
For $\ve \in (0,\sqrt{3}/4]$, $\g_{n,\ve}(3/2) \rightarrow 2$ in
probability as $n \rightarrow \infty$, and for $\ve \in
(\sqrt{3}/4,\sqrt{3}/3)$, $\g_{n,\ve}(3/2) \rightarrow 1$ in
probability as $n \rightarrow \infty$.

Similarly, the stochastic ordering result of Theorem 6 holds for
association for all $\ve$ and $n<\infty$, with the inequalities
being reversed.

Notice that under segregation with $\ve \in (0,\sqrt{3}/4)$,
$\g_{n,\ve}(r)$ is degenerate in the limit except for
$r=(3-\sqrt{3}\,\ve)/2 $. With $\ve \in (\sqrt{3}/4,\sqrt{3}/3)$,
$\g_{n,\ve}(r)$ is degenerate in the limit except for
$r=\sqrt{3}/\ve-2$. Under association with $\ve \in (0,\sqrt{3}/4)$,
$\g_{n,\ve}(r)$ is degenerate in the limit except for
$r=\frac{3}{2\,(1-\sqrt{3}\,\ve)}$.

The mean domination number of the proximity catch digraph,
$\overline G_J:=\frac{1}{J}\,\sum_{j=1}^J \g_{n_j}(3/2)$, is a test
statistic for the segregation/association alternative; rejecting for
extreme values of $\overline G_J$ is appropriate, since under
segregation we expect $\overline G_J$ to be small, while under
association we expect $\overline G_J$ to be large. Using the
equivalent test statistic
\begin{equation}
S = \sqrt{J} (\overline G_J - \mu)/\sigma,
\end{equation}
the asymptotic critical value for the one-sided level $\alpha$ test
against segregation is given by
\begin{equation}
z_{1-\alpha} = \Phi^{-1}(\alpha)
\end{equation}
where $\Phi(\cdot)$ is the standard normal distribution function.
The test rejects for $S<z_{1-\alpha}$. Against association, the test
rejects for $S>z_{\alpha}$.

\begin{figure}[ht]
\centering \epsfig{figure=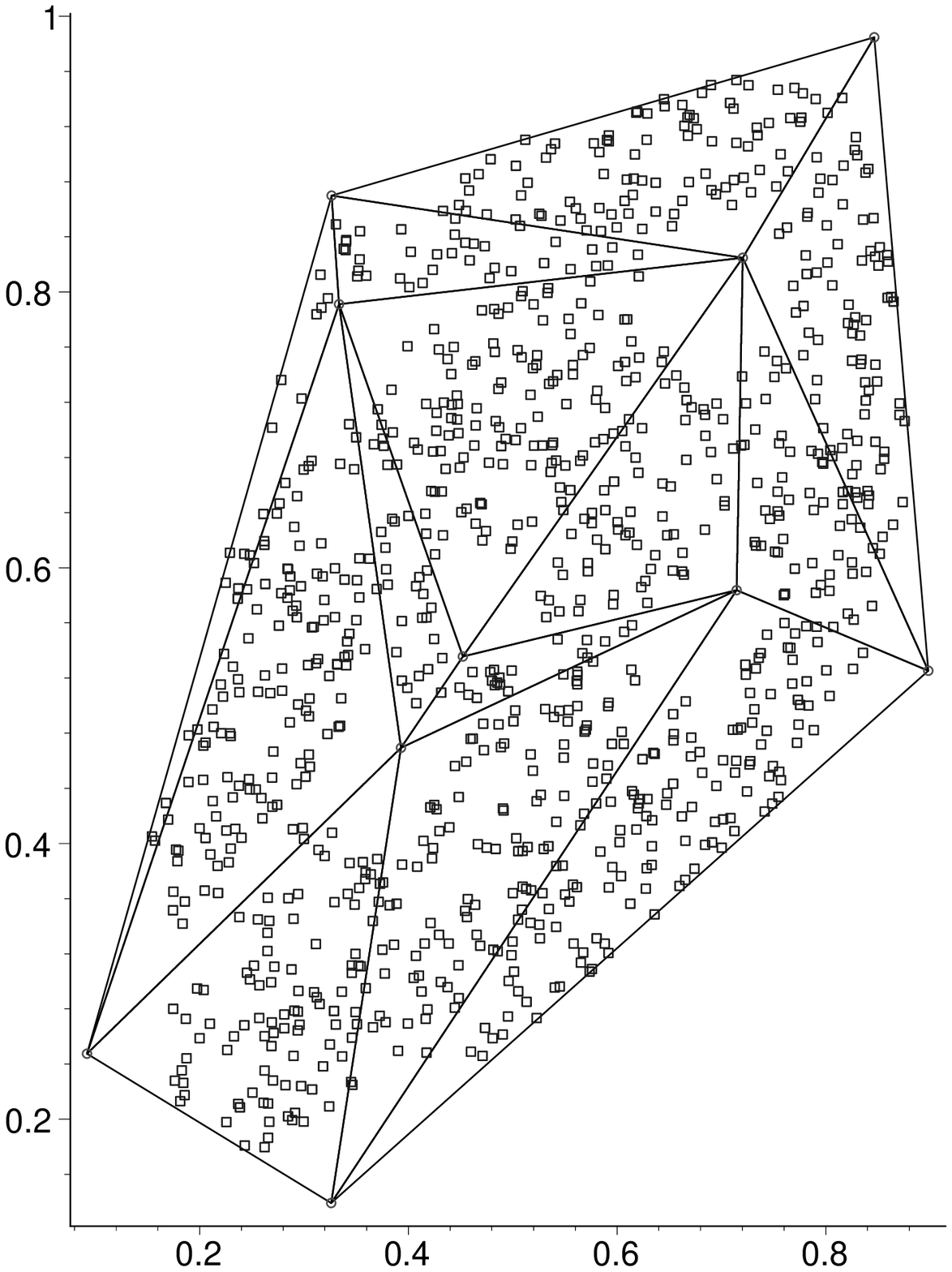, height=170pt,
width=170pt} \epsfig{figure=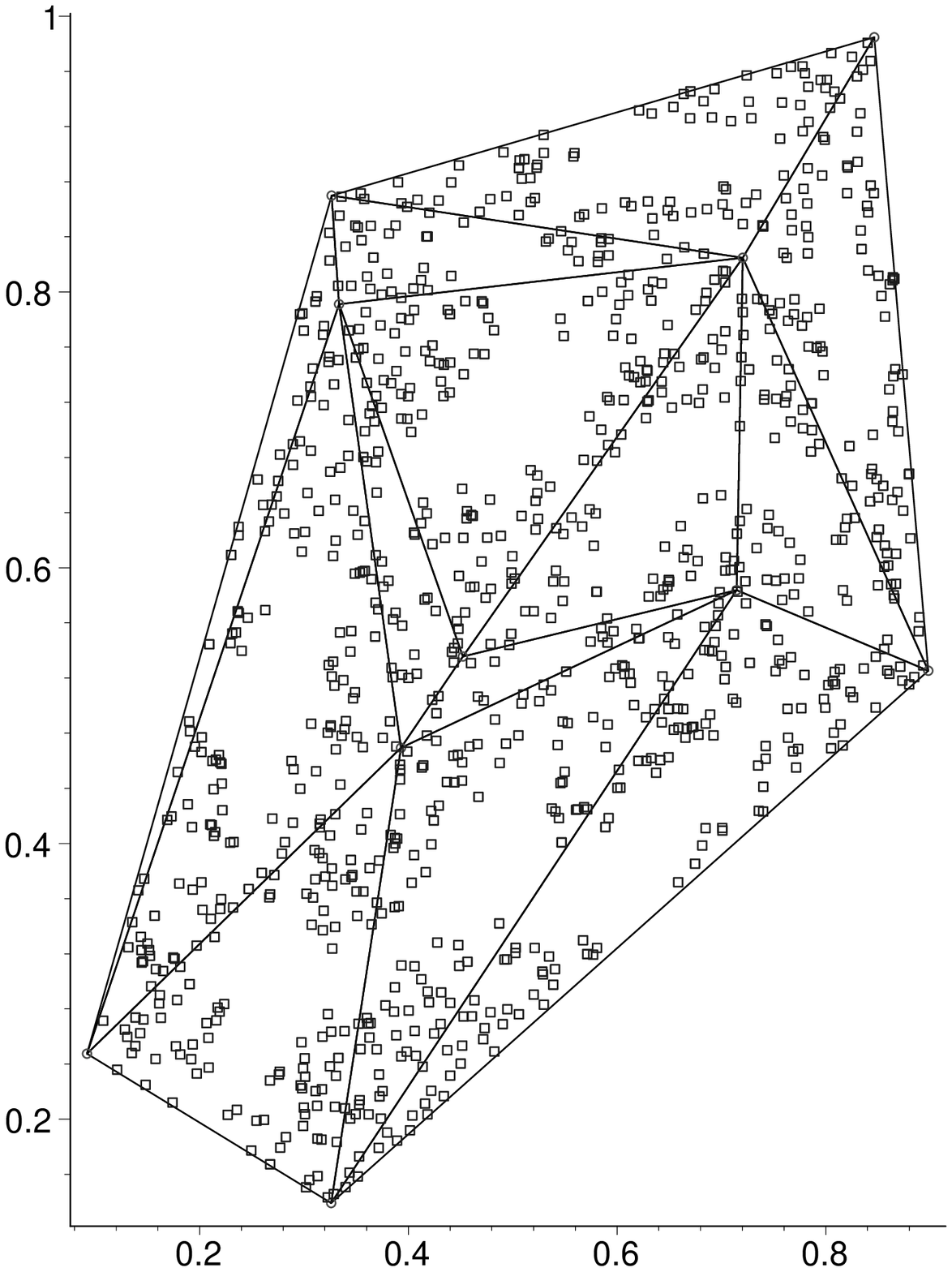,
height=170pt, width=170pt} \caption{ \label{fig:deldata} A
realization of segregation (left) and association (right) for
$|\Y|=10$, $J=13$, and $n=1000$. }
\end{figure}

Depicted in Figure \ref{fig:deldata} are the segregation with
$\delta=1/16$ and association with $\delta=1/4$ realizations for
$|\Y|=10$ and $J=13$, and $n=1000$. The associated mean domination
numbers are $2.308,\;1.923$, and $3.000$, for the null realization
in Figure \ref{fig:deldatacsr} and the segregation and association
alternatives in Figure \ref{fig:deldata}, respectively, yielding
p-values $.660,\;.003$ and $0.000$. We also present a Monte Carlo
power investigation in Section 6 for these cases.

{\bf Theorem 7}: (Consistency) Let $J^*(\alpha,\ve):=\left \lceil
\Bigl(\frac{\sigma \cdot z_{\alpha}}{\mu-\overline G_J} \Bigr)^2
\right \rceil$ where $\lceil \cdot \rceil$ is the ceiling function
and $\ve$-dependence is through $\overline G_J$ under a given
alternative. Then the test against $H^S_{\ve}$ which rejects for $S
< z_{1-\alpha}$ is consistent for all $\ve \in (0,\sqrt{3}/3)$ and
$J \ge J^*(1-\alpha,\ve)$, and the test against $H^A_{\ve}$ which
rejects for $S>z_{\alpha}$ is consistent for all $\ve \in
(0,\sqrt{3}/3)$ and $J \ge J^*(\alpha,\ve)$.

{\bf Proof:} Let $\ve>0$. Under $H^S_{\ve}$, $\g_{n,\ve}(3/2)$ is
degenerate in the limit as $n \rightarrow \infty$, which implies
$\overline G_J$ is a constant a.s. In particular, for $\ve \in
(0,\sqrt{3}/4]$, $\overline G_J=2$ and for $\ve \in
(\sqrt{3}/4,\sqrt{3}/3)$, $\overline G_J=1$ a.s. as $n \rightarrow
\infty$. Then the test statistic $S = \sqrt{J} (\overline G_J -
\mu)/\sigma$ is a constant a.s. and $J \ge J^*(1-\alpha,\ve)$
implies that $S< z_{1-\alpha}$ a.s. Hence consistency follows for
segregation.

Under $H^A_{\ve}$, as $n \rightarrow \infty$, $\overline G_J=3$ for
all $\ve \in (0,\sqrt{3}/3)$, a.s. Then $J \ge J^*(\alpha,\ve)$
implies that $S > z_{\alpha}$ a.s., hence consistency follows for
association. $\blacksquare$

\section{Monte Carlo Power Analysis}
\begin{figure}[ht]
\centering \psfrag{kernel density estimate}{ \Huge{\bf{kernel
density estimate}}} \psfrag{relative density}{ \Huge{\bf{relative
density}}} \rotatebox{-90}{ \resizebox{2.1 in}{!}{
\includegraphics{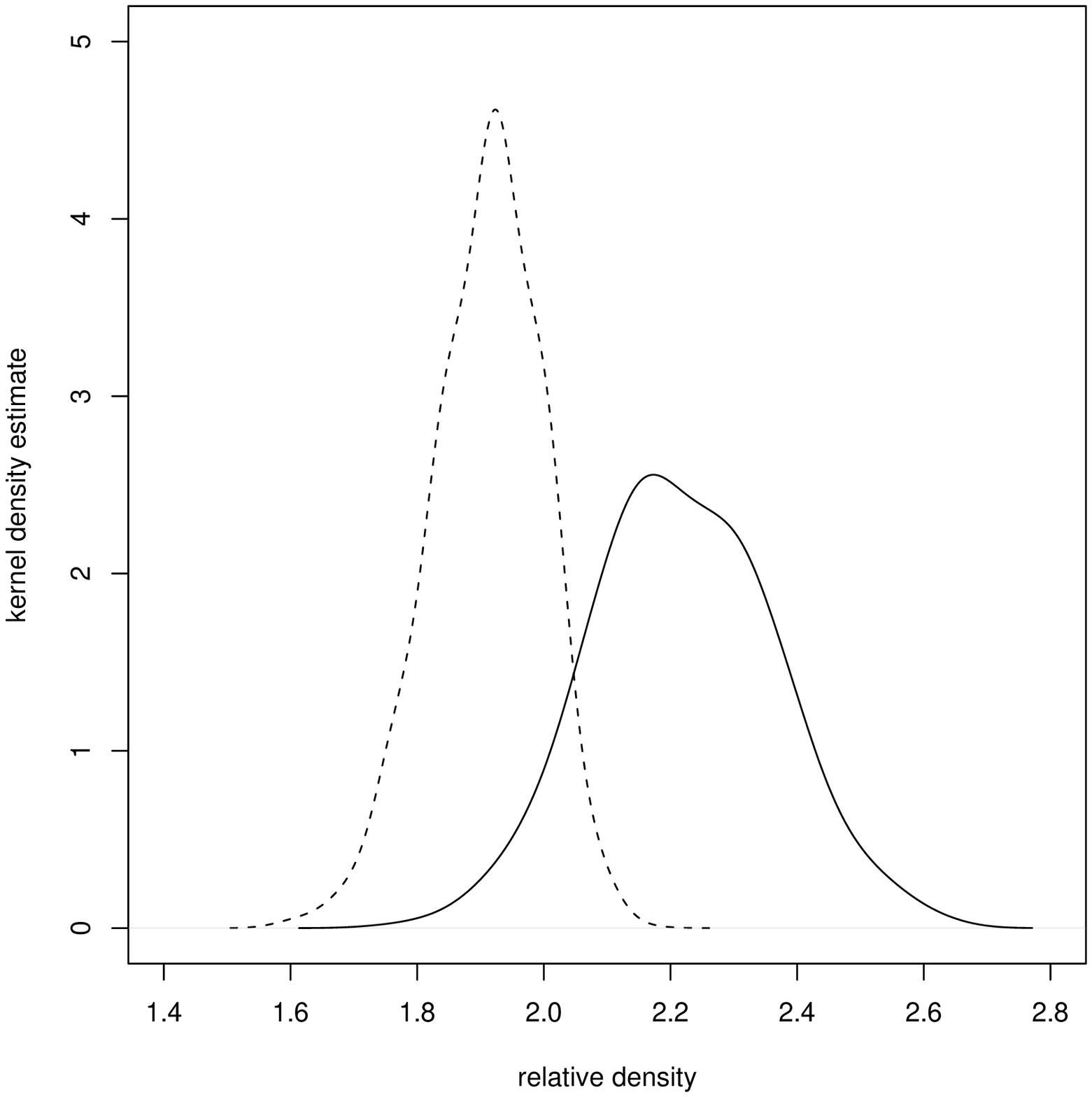} } } \rotatebox{-90}{
\resizebox{2.1 in}{!}{ \includegraphics{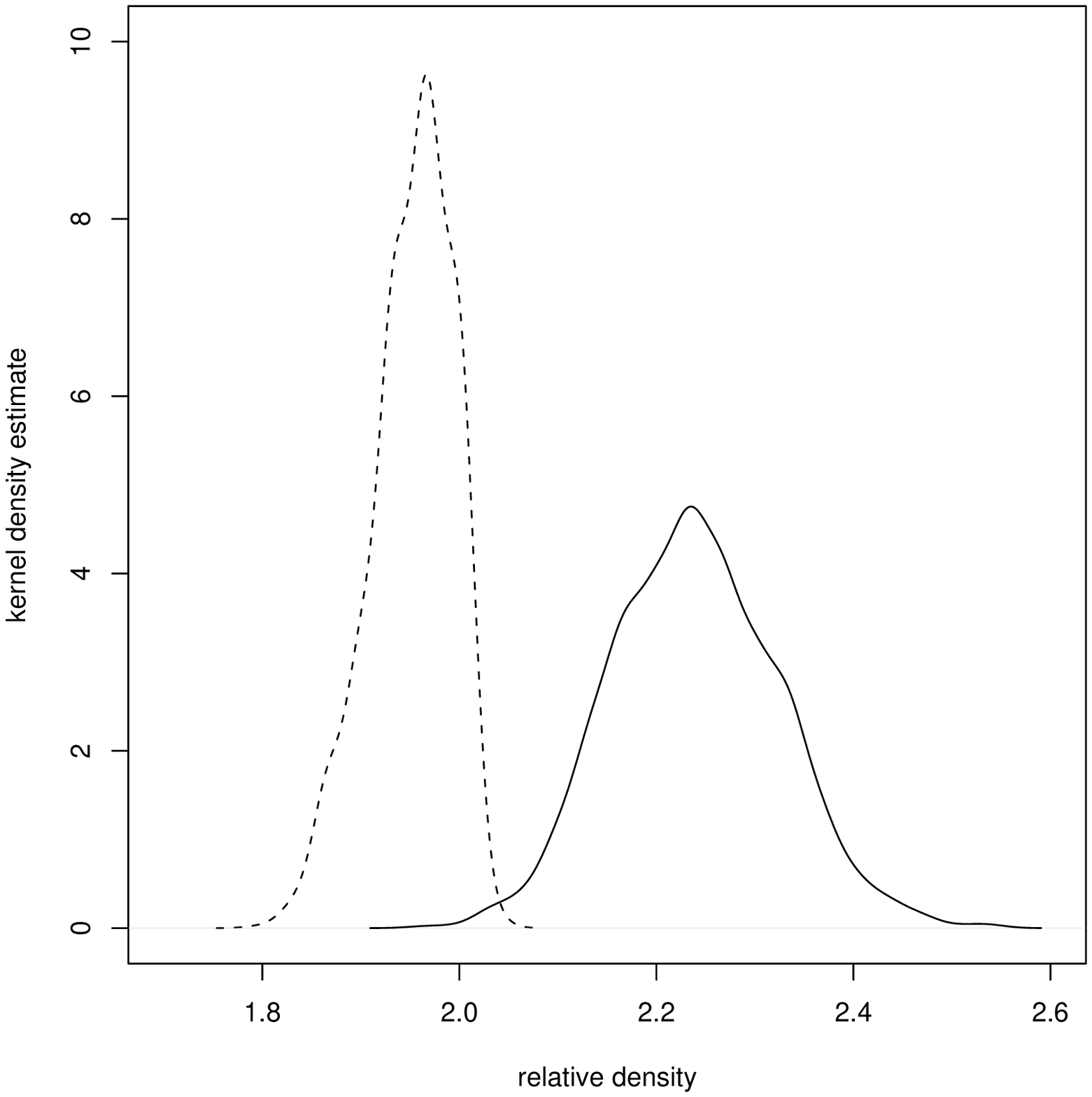} } }
\caption{ \label{fig:CSRvsSeg} Two Monte Carlo experiments against
the segregation alternatives $H^S_{\sqrt{3}/8}$ with $\delta=1/16$.
Depicted are kernel density estimates of $\overline G_J$ for $J=13$
and $n=1000$ with $1000$ replicates (left) and $J=30$ and $n=5000$
with $1000$ replicates (right) under the null (solid) and
alternative (dashed). }
\end{figure}

In Figure \ref{fig:CSRvsSeg}, we observe empirically that even under
mild segregation we obtain considerable separation between the
kernel density estimates under null and segregation cases for
moderate $J$ and $n$ values suggesting high power at $\alpha=.05$. A
similar result is observed for association. With $J=13$ and
$n=1000$, under $H_0$, the estimated significance level is
$\widehat{\alpha}=.09$ relative to segregation, and
$\widehat{\alpha}=.07$ relative to association. Under
$H^S_{\sqrt{3}/8}$, the empirical power (using the asymptotic
critical value) is $\widehat{\beta}=.97$, and under
$H^A_{\sqrt{3}/21}$, $\widehat{\beta}=1.00$. With $J=30$ and
$n=5000$, under $H_0$, the estimated significance level is
$\widehat{\alpha}=.06$ relative to segregation, and
$\widehat{\alpha}=.04$ relative to association. The empirical power
is $\widehat{\beta}=1.00$ for both alternatives.

We also estimate the empirical power by using the empirical critical
values. With $J=13$ and $n=1000$, under $H^S_{\sqrt{3}/8}$, the
empirical power is $\widehat{\beta}_{mc}=.72$ at empirical level
$\widehat{\alpha}_{mc}=.033$ and under $H^A_{\sqrt{3}/21}$ the
empirical power is $\widehat{\beta}_{mc}=1.00$ at empirical level
$\widehat{\alpha}_{mc}=.03$. With $J=30$ and $n=5000$, under
$H^S_{\sqrt{3}/8}$, the empirical power is
$\widehat{\beta}_{mc}=1.00$ at empirical level
$\widehat{\alpha}_{mc}=.034$ and under $H^A_{\sqrt{3}/21}$ the
empirical power is $\widehat{\beta}_{mc}=1.00$ at empirical level
$\widehat{\alpha}_{mc}=.04$.

\section{Extension to Higher Dimensions}
The extension to $\R^d$ for $d > 2$ is straightforward. Let $\Y =
\{\y_1,\y_2,\cdots,\y_{d+1}\}$ be $d+1$ non-coplanar points. Denote
the simplex formed by these $d+1$ points as $\mathfrak S (\Y)$. (A
simplex is the simplest polytope in $\R^d$ having $d+1$ vertices,
$d\,(d+1)/2$ edges, and $d+1$ faces of dimension $(d-1)$.) For $r
\in [1,\infty]$, define the $r$-factor proximity map as follows.
Given a point $x$ in $\mathfrak S (\Y)$, let $\y := \arg\min_{\y \in
\Y} \mbox{volume}(Q_{\y}(x))$ where $Q_{\y}(x)$ is the polytope with
vertices being the $d\,(d+1)/2$ midpoints of the edges, the vertex
$\y$ and $x$. That is, the vertex region for vertex $v$ is the
polytope with vertices given by $v$ and the midpoints of the edges.
Let $v(x)$ be the vertex in whose region $x$ falls. If $x$ falls on
the boundary of two vertex regions, we assign $v(x)$ arbitrarily.
Let $\varphi(x)$ be the face opposite to vertex $v(x)$, and
$\eta(v(x),x)$ be the hyperplane parallel to $\varphi(x)$ which
contains $x$. Let $d(v(x),\eta(v(x),x))$ be the (perpendicular)
Euclidean distance from $v(x)$ to $\eta(v(x),x)$. For $r \in
[1,\infty)$, let $\eta_r(v(x),x)$ be the hyperplane parallel to
$\varphi(x)$ such that
$d(v(x),\eta_r(v(x),x))=r\,d(v(x),\eta(v(x),x))$ and
$d(\eta(v(x),x),\eta_r(v(x),x))< d(v(x),\eta_r(v(x),x))$. Let
$\mathfrak S_r(x)$ be the polytope similar to and with the same
orientation as $\mathfrak S$ having $v(x)$ as a vertex and
$\eta_r(v(x),x)$ as the opposite face. Then the $r$-factor proximity
region $\NY^r(x):=\mathfrak S_r(x) \cap \mathfrak S(\Y)$.
Also, let $\zeta_j(x)$ be the hyperplane such that $\zeta_j(x) \cap \mathfrak S(\Y) \not=\emptyset$ and $r\,d(\y_j,\zeta_j(x))=d(\y_j,\eta(\y_j,x))$ for $j=1,2,\ldots,d+1$.  Then $\G^r_1(x)=\cup_{j=1}^{d+1} (\G^r_1(x)\cap R(\y_j))$ where $\G^r_1(x)\cap R(\y_j)=\{z \in R(\y_j): d(\y_j,\eta(\y_j,z)) \ge d(\y_j,\zeta_j(x)\}$, for $j=1,2,3$. 

Theorem 1 generalizes, so that any simplex $\mathfrak S$ in $\R^d$
can be transformed into a regular polytope (with egdes being equal
in length and faces being equal in volume) preserving uniformity.
Delaunay triangulation becomes Delaunay tessellation in $\R^d$,
provided that no more than $d+1$ points being cospherical (lying on
the boundary of the same sphere). In particular, with $d=3$, the
general simplex is a tetrahedron (4 vertices, 4 triangular faces and
6 edges), which can be mapped into a regular tetrahedron (4 faces
are equilateral triangles) with vertices
$(0,0,0),\,(1,0,0)\,(1/2,\sqrt{3}/2,0),\,(1/2,\sqrt{3}/6,\sqrt{6}/3)$.
Let $\g_n(r,d)$ be the domination number for the extension to
$\R^d$.  Then it is easy to see that $\g_n(r,3)$ is nondegenerate as
$n \rightarrow \infty$ for $r=4/3$, and otherwise degenerate. In
$\R^d$, it can be seen that $\g_n(r,d)$ is nondegenerate in the
limit only for $r=(d+1)/d$.  Moreover, it can be shown that $\lim_{n
\rightarrow \infty}P(2 \le \g_n((d+1)/d,d) \le d+1)=1$, and we
conjecture that
$$\lim_{n \rightarrow \infty}P(d \le \g_n((d+1)/d,d) \le d+1)=1.$$

\subsection{Discussion}
\label{sec:discussion} In this article we investigate the
mathematical properties of a domination number method for the
analysis of spatial point patterns.

The first proximity map related to $r$-factor proximity map,
$\NY^r$, in literature is the \emph{spherical proximity map},
$N_S(x):=B(x,r(x))$, (which is called CCCD in the literature, see
\cite{priebe:2001}, \cite{devinney:2002a}, \cite{marchette:2003},
\cite{priebe:2003b}, and \cite{priebe:2003a}). A slight variation of
$N_S$ is the \emph{arc-slice proximity map} $N_{AS}(x):=B(x,r(x))
\cap T(x)$ where $T(x)$ is the Delaunay cell that contains $x$ (see
\cite{ceyhan:2003a}). Furthermore, Ceyhan and Priebe introduced the
central similarity proximity map, $N_{CS}$, in \cite{ceyhan:2003a}.
The $r$-factor proximity map, when compared to the others, has the
advantages that the asymptotic distribution of the domination number
$\g_n(r)$ is tractable (see Theorem 3). The distribution of the
domination number of the proximity catch digraphs based on $N_S$ or
$N_{AS}$ is not tractable, and that of $N_{CS}$ is an open problem.
Furthermore, $N_{CS}$ and $\NY^r$ enjoy the geometry invariance
property over triangles for uniform data. Moreover, while finding
the exact minimum dominating sets is an NP-Hard problem for $N_S$,
$N_{AS}$, and $N^{\tau}_{CS}$, the exact minimum dominating sets can
be found in polynomial time for $N_\Y^r$. Additionally, $N_{AS}(x)$,
$N_\Y^r(x)$, and $N^{\tau}_{CS}(x)$ are well defined only for $x \in
C_H(\Y)$, the convex hull of $\Y$, whereas $N_S(x)$ is well defined
for all $x \in \R^d$.

The $N_S$ (the proximity map associated with CCCD) is used in
classification in the literature, but not for testing spatial
patterns between two or more classes. We develop a technique to test
the patterns of segregation or association. There are many tests
available for segregation and association in ecology literature. See
\cite{dixon:1994} for a survey on these tests and relevant
references. Two of the most commonly used tests are Pielou's
$\chi^2$ test of independence and Ripley's test based on $K(t)$ and
$L(t)$ functions. However, the test we introduce here is not
comparable to either of them. Our method deals with a slightly
different type of data than most methods to examine spatial
patterns. The sample size for one type of point (type $\X$ points)
is much larger compared to the the other (type $\Y$ points).

The null hypothesis we consider is considerably more restrictive
than current approaches, which can be used much more generally. The
null hypothesis for testing segregation or association can be
described in two slightly different forms (\cite{dixon:1994}):
\begin{itemize}
\item[(i)]complete spatial randomness,
that is, each class is distributed randomly throughout the area of
interest. It describes both the arrangement of the locations and the
association between classes.
\item[(ii)] random labeling of locations, which is less restrictive than spatial randomness,
in the sense that arrangement of the locations can either be random
or non-random.
\end{itemize}
Our test is closer to the former in this regard.

%

\section{Appendix}
\subsection*{Proof of Proposition 1}

To prove Proposition 1, we show that the expected locus of the
boundary of the $\G_1$-region, $\partial(\G_1^r(\X_n))$, goes to
$\partial(\varsigma_{\Y}^r)$ as $n \rightarrow \infty$ by showing
that the expected loci of $X_{e_j}$ are $e_j$ for $j=1,2,3$. See
\cite{ceyhan:2003b} for the details.

For sufficiently large $n$ and given $X_{e_j}=(x_j,y_j)$ for
$j=1,2,3$,
$$A(\G_1^{3/2}(\X_n)) = \sqrt {3}/9(3\,{x_2}^2-6\,x_2+2\,\sqrt {3}y_2\,x_2-2\,\sqrt {3}y_2+{y_2}^2+3+{y_3}^2-2\,\sqrt {3}y_3\,x_3+3\,{x_3}^2+4\,{y_1}^2).$$
The asymptotically accurate joint pdf of $X_{e_j}$'s is
$$f_3(\zeta) = n(n-1)(n-2) \bigl(\sqrt {3}/36(-2\,\sqrt {3}y_1+\sqrt {3}y_3-3\,x_3+\sqrt {3}y_2+3\,x_2)^2\bigr) ^{n-3}/(\sqrt{3}/4)^n $$
with the support $D_S=\{\zeta=(x_1,y_1,x_2,y_2,x_3,y_3) \in \R^6:\;
(x_j,y_j) \text{'s are distinct} \}.$ Then for sufficiently large
$n$, $\E[A(\G_1^{3/2}(\X_n))] \approx \int_{D_S}
A(\G_1^{3/2}(\X_n))f_3(\zeta)d\zeta,$ which goes to $0$ as $n
\rightarrow \infty$ at the rate $O(n^{-2})$. See \cite{ceyhan:2003b}
for the details.

\subsection*{Proof of Theorem 3}
We know that $\g_n(r) \le 3$ a.s. for all $r \in [1,\infty]$ and all
$n$. First we show that $\lim_{n \rightarrow
\infty}P(\g_n(3/2)>1)=1$.

Note that $P(\g_n(3/2) > 1)=P(\X_n \cap \G_1^{3/2}(\X_n)
=\emptyset)$. Then we find $P(\X_n \cap \G_1^{3/2}(\X_n)
=\emptyset,\,E_{2}(n,\ve))$ where $E_2(n,\ve)$ is the event such
that $\frac{2\ve }{\sqrt{3}} \le X_1 \le 1-\frac{2\ve}{\sqrt{3}}$
and $0 \le Z_1 \le \ve$, and $1/2 \le X_2 \le
1-\frac{2\ve}{\sqrt{3}}$, $\sqrt{3}(1-X_2)-\ve \le Z_2 \le
\sqrt{3}(1-X_2)$, and $\frac{2\ve }{\sqrt{3}} \le X_3 \le 1/2$, and
$\sqrt{3}\,X_3-\ve \le Z_3 \le \sqrt{3}\,X_3$. First letting $n
\rightarrow \infty$, then $\ve \rightarrow 0$,  yields the desired
result. See \cite{ceyhan:2003b} for the details.

Next, $\lim_{n \rightarrow \infty}P(\g_n(3/2) \le 2)=\lim_{n
\rightarrow \infty}P(\g_n(3/2)=2)$, since $\lim_{n \rightarrow
\infty}P(\g_n(3/2)=1)=0$. Let
$$Q_j:=\argmin_{x \in \X_n \cap R(\y_j)} d(x,e_j)=\argmax_{x \in \X_n \cap R(\y_j)} d(\ell(\y_j,x),e_j)$$
 where $e_j$ is the edge opposite vertex $\y_j$ for $j=1,2,3$ and let $q_j=(x_j,y_j)$ be the realization of $Q_j$ for $j=1,2,3$. Then $\g_n(3/2)\le 2$ iff $\X_n \subset N_{\Y}^{3/2}(Q_1) \cup N_{\Y}^{3/2}(Q_2)$ or $\X_n \subset N_{\Y}^{3/2}(Q_1) \cup N_{\Y}^{3/2}(Q_3)$ or $\X_n \subset N_{\Y}^{3/2}(Q_2) \cup N_{\Y}^{3/2}(Q_3)$.

Let the events $E_{i,j}:=\X_n \subset N_{\Y}^{3/2}(Q_i) \cup
N_{\Y}^{3/2}(Q_j)$ for $(i,j)=\{(1,2),(1,3),(2,3)\}$.  Then
$$P(\g_n(3/2)\le 2)=P(E_{1,2})+P(E_{1,3})+P(E_{2,3})-P(E_{1,2}\cap E_{1,3})-P(E_{1,2}\cap E_{2,3})-P(E_{1,3}\cap E_{2,3})+P(E_{1,2}\cap E_{1,3} \cap E_{2,3}).$$
By symmetry, $P(E_{1,2})=P(E_{1,3})=P(E_{2,3})$ and $P(E_{1,2}\cap
E_{1,3})=P(E_{1,2}\cap E_{2,3})=P(E_{1,3}\cap E_{2,3})$. Hence
$$P(\g_n(3/2)\le 2)=3\,\Bigl[P(E_{1,2})-P(E_{1,2}\cap E_{1,3})\Bigr]+P(E_{1,2}\cap E_{1,3} \cap E_{2,3}).$$

We find $P(E_{1,2})$, by finding the asymptotically accurate joint
pdf of $Q_1,\,Q_2$. Let $T(Q_j)$ be the triangle formed by the
median lines at $\y_k$ and $\y_l$ for $k,l\not=j$ and
$\ell(\y_j,Q_j)$, and let $\ve>0$ be small enough such that $T(Q_j)
\subset R(\y_j)$, for $j=1,2,3$. Then the asymptotically accurate
joint pdf of $Q_1,\,Q_2$ is
$$f_{1,2}(x_1,y_1,x_2,y_2)=n\,(n-1)\,\frac{1}{A(T(\Y))^2}\left(\frac{A(T(\Y))-A(T(q_1))-A(T(q_2))}{A(T(\Y))} \right)^{n-2}$$
 where $A(T(q_1))=\sqrt {3}/36 \left(-2\,\sqrt {3}+3\,y_1+3\,\sqrt {3}x_1 \right)^2$ and $A(T(q_2))=\sqrt {3}/36 \left(-3\,y_2-\sqrt {3}+3\,\sqrt {3}x_2 \right)^2$ with domain $D_I=\{(x_1,y_1) \in R(\y_1):\; y_1 \ge -\frac{\sqrt {3}}{3}\,+\sqrt {3}x_1+\sqrt {3}\ve,\;(x_2,y_2) \in R(\y_2):\; y_2 \le -\frac{\sqrt {3}}{3}\,+\sqrt {3}x_2-\sqrt {3}\,\ve\}$ with $\ve>0$ be small enough such that $T(Q_j) \subset R(\y_j)$, for $j=1,2,3$.

Then $P(E_{1,2})\approx .4126$ (which is found numerically). See
\cite{ceyhan:2003b} for the details.

Similarly we find $P(E_{1,2} \cap E_{1,3})$, by finding the joint
pdf of $Q_1,\,Q_2,\,Q_3$,  where $T(q_3)$ is the triangle with
vertices $\frac{1}{3}\,(\sqrt {3}-3\,y_3)\sqrt
{3},y_3),\,(1/2,\sqrt{3}/6),(\sqrt {3}y_3,y_3)$. Then the
asymptotically accurate joint pdf of $Q_1,\,Q_2,\,Q_3$ is
$$f_{123}(x_1,y_1,x_2,y_2,x_3,y_3)=n\,(n-1)\,(n-2)\,\frac{1}{A(T(\Y))^3}\left(\frac{A(T(\Y))-A(T(q_1))-A(T(q_2))-A(T(q_3))}{A(T(\Y))}\right)^{n-3}$$
where $A(T(q_3))=\frac{\sqrt {3}}{36}\,(-\sqrt {3}+6\,y_3)^2$ with
domain $D_I=\{(x_1,y_1) \in R(\y_1):\; y_1 \ge -\frac{\sqrt
{3}}{3}\,+\sqrt {3}x_1+\sqrt {3}\,\ve,\;(x_2,y_2) \in R(\y_2):\; y_2
\ge -\frac{\sqrt {3}}{3}\,+\sqrt {3}x_2-\sqrt {3}\,\ve,\;(x_3,y_3)
\in R(\y_3):\; y_3 \le \frac{\sqrt{3}}{6}+\ve\}$.

Then $P(E_{1,2} \cap E_{1,3}) \approx .2009$ (see
\cite{ceyhan:2003b} for the details.)

Likewise, we find $P(E_{1,2} \cap E_{1,3}\cap E_{2,3})\approx .1062$
(see \cite{ceyhan:2003b} for the details.)

Hence we get $\lim_{n \rightarrow \infty}P(\gamma(\X_n,
N_{\Y}^{3/2})=2) \approx .7413$, and $\lim_{n \rightarrow
\infty}P(\gamma(\X_n, N_{\Y}^{3/2})=3) \approx .2587$.

\end{document}

%% file: Nofnu2.pstex_t
\begin{picture}(0,0)%
\includegraphics{Nofnu2.pstex}%
\end{picture}%
\setlength{\unitlength}{3947sp}%
\begingroup\makeatletter\ifx\SetFigFont\undefined%
\gdef\SetFigFont#1#2#3#4#5{%
  \reset@font\fontsize{#1}{#2pt}%
  \fontfamily{#3}\fontseries{#4}\fontshape{#5}%
  \selectfont}%
\fi\endgroup%
\begin{picture}(11231,8043)(64,-7348)
\put(1051,-6511){\makebox(0,0)[lb]{\smash{\SetFigFont{22}{14.4}{\rmdefault}{\mddefault}{\updefault}{\color[rgb]{0,0,0}$\y_1=v(x)$}%
}}}
\put(3376,-4936){\makebox(0,0)[lb]{\smash{\SetFigFont{22}{14.4}{\rmdefault}{\mddefault}{\updefault}{\color[rgb]{0,0,0}$x$}%
}}}
\put(5326,-3736){\makebox(0,0)[lb]{\smash{\SetFigFont{22}{14.4}{\rmdefault}{\mddefault}{\updefault}{\color[rgb]{0,0,0}$M_C$}%
}}}
\put(751,-1861){\rotatebox{320.0}{\makebox(0,0)[lb]{\smash{\SetFigFont{22}{14.4}{\rmdefault}{\mddefault}{\updefault}{\color[rgb]{0,0,0}$\ell(v(x),x)$}%
}}}}
\put(2992,143){\rotatebox{315.0}{\makebox(0,0)[lb]{\smash{\SetFigFont{22}{14.4}{\rmdefault}{\mddefault}{\updefault}{\color[rgb]{0,0,0}$\ell_2(v(x),x)$}%
}}}}
\put(4876,539){\makebox(0,0)[lb]{\smash{\SetFigFont{22}{14.4}{\rmdefault}{\mddefault}{\updefault}{\color[rgb]{0,0,0}$\y_3$}%
}}}
\put(6975,-1389){\rotatebox{310.0}{\makebox(0,0)[lb]{\smash{\SetFigFont{22}{14.4}{\rmdefault}{\mddefault}{\updefault}{\color[rgb]{0,0,0}$e(x)$}%
}}}}
\put(11251,-6136){\makebox(0,0)[lb]{\smash{\SetFigFont{22}{14.4}{\rmdefault}{\mddefault}{\updefault}{\color[rgb]{0,0,0}$\y_2$}%
}}}
\put(2926,-7186){\rotatebox{45.0}{\makebox(0,0)[lb]{\smash{\SetFigFont{22}{14.4}{\rmdefault}{\mddefault}{\updefault}{\color[rgb]{0,0,0}$d(v(x),\ell_2(v(x),x))=2\,d(v(x),\ell(v(x),x))$}%
}}}}
\put(960,-5299){\rotatebox{45.0}{\makebox(0,0)[lb]{\smash{\SetFigFont{22}{14.4}{\rmdefault}{\mddefault}{\updefault}{\color[rgb]{0,0,0} $d(v(x),\ell(v(x),x))$}%
}}}}
\end{picture}

%% file: Gammaofnu.pstex_t
\begin{picture}(0,0)%
\includegraphics{Gammaofnu.pstex}%
\end{picture}%
\setlength{\unitlength}{3947sp}%
\begingroup\makeatletter\ifx\SetFigFont\undefined%
\gdef\SetFigFont#1#2#3#4#5{%
  \reset@font\fontsize{#1}{#2pt}%
  \fontfamily{#3}\fontseries{#4}\fontshape{#5}%
  \selectfont}%
\fi\endgroup%
\begin{picture}(10684,8007)(589,-7348)
\put(1051,-6511){\makebox(0,0)[lb]{\smash{{\SetFigFont{22}{14.4}{\rmdefault}{\mddefault}{\updefault}{\color[rgb]{0,0,0}$\y_1$}%
}}}}
\put(4876,539){\makebox(0,0)[lb]{\smash{{\SetFigFont{22}{14.4}{\rmdefault}{\mddefault}{\updefault}{\color[rgb]{0,0,0}$\y_3$}%
}}}}
\put(1405,-2508){\rotatebox{320.0}{\makebox(0,0)[lb]{\smash{{\SetFigFont{22}{14.4}{\rmdefault}{\mddefault}{\updefault}{\color[rgb]{0,0,0}$\ell(\y_1,x)$}%
}}}}}
\put(3376,-4936){\makebox(0,0)[lb]{\smash{{\SetFigFont{22}{14.4}{\rmdefault}{\mddefault}{\updefault}{\color[rgb]{0,0,0}$x$}%
}}}}
\put(2626,-2011){\makebox(0,0)[lb]{\smash{{\SetFigFont{22}{14.4}{\rmdefault}{\mddefault}{\updefault}{\color[rgb]{0,0,0}$\xi_3(2,x)$}%
}}}}
\put(7533,-5107){\rotatebox{65.0}{\makebox(0,0)[lb]{\smash{{\SetFigFont{22}{14.4}{\rmdefault}{\mddefault}{\updefault}{\color[rgb]{0,0,0}$\xi_2(2,x)$}%
}}}}}
\put(960,-5299){\rotatebox{45.0}{\makebox(0,0)[lb]{\smash{{\SetFigFont{22}{14.4}{\rmdefault}{\mddefault}{\updefault}{\color[rgb]{0,0,0}$d(\y_1,\ell(\y_1,x))=r\,d(\y_1,\xi_1(2,x))$}%
}}}}}
\put(2026,-7036){\rotatebox{45.0}{\makebox(0,0)[lb]{\smash{{\SetFigFont{22}{14.4}{\rmdefault}{\mddefault}{\updefault}{\color[rgb]{0,0,0}$d(\y_1,\xi_1(2,x))$}%
}}}}}
\put(3867,-6839){\rotatebox{320.0}{\makebox(0,0)[lb]{\smash{{\SetFigFont{22}{14.4}{\rmdefault}{\mddefault}{\updefault}{\color[rgb]{0,0,0}$\xi_1(2,x)$}%
}}}}}
\put(11026,-6436){\makebox(0,0)[lb]{\smash{{\SetFigFont{22}{14.4}{\rmdefault}{\mddefault}{\updefault}{\color[rgb]{0,0,0}$\y_2$}%
}}}}
\put(5626,-4036){\makebox(0,0)[lb]{\smash{{\SetFigFont{22}{14.4}{\rmdefault}{\mddefault}{\updefault}{\color[rgb]{0,0,0}$M_C$}%
}}}}
\end{picture}%